# Decoupling Periodic Systems: An Algebraic Approach

Vladimír Kučera, *Life Member, IEEE*

***Abstract*— This paper addresses the problem of row-by-row (or diagonal) decoupling of discrete-time linear multi-input multi-output systems with periodic time-varying coefficients using periodic state feedback. Previous solutions have tackled row-by-row decoupling using dynamic compensation for square systems and block-decoupling through regular state feedback for nonsquare systems with more outputs than inputs. While it appears likely that a row-by-row state feedback solution for square systems can be deduced from these findings, a direct argument seems more appropriate here as it presents a natural extension for decoupling nonsquare systems with more inputs than outputs. This extension, which necessitates nonregular state feedback, has yet to be explored for periodic systems. Our approach is purely algebraic, based on a time-invariant representation of the periodic system.**

## I. Introduction

Decoupling, also known as noninteractive control is a recognized issue in control theory. It consists of compensating a given multi-input multi-output system so that each system output is controlled by one control input without being influenced by the other inputs. The compensation scheme is generally taken as state feedback, and the decoupled system is required to be stable.

For linear time-invariant systems, the problem was rigorously posed in [1] and then solved in [2] for square systems, which have the same number of inputs and outputs, using static-state feedback. More general solutions were reported in [3] using the vector space approach, and still further insight was provided in [4] and [5] by applying the transfer function approach. Despite long-time efforts, a general solution for linear time-invariant systems with more inputs than outputs has recently been achieved [6]. Systems with more outputs than inputs cannot be diagonally decoupled; only block decoupling is possible.

An alternative compensation scheme consists of output feedback, either static or dynamic. The ultimate solution to this problem was presented in [7]. While output feedback is an appealing alternative from the implementation point of view, some systems can be decoupled by static-state feedback but cannot be decoupled by output feedback in any form.

Linear systems with periodically varying coefficients are an intermediate class between time-invariant and general time-varying systems, which significantly enlarges the variety of processes that can be modeled. For discrete-time periodic systems, in particular, various modifications of the decoupling problem were solved under restrictive hypotheses about the system and the decoupling feedback. B*lock-decoupling* by regular state feedback was studied in [8] and the row-by-row decoupling of square systems by *dynamic compensations* was considered in [9].

Control problems for *discrete-time* periodic systems, can be equivalently recast as problems for time-invariant systems by exploiting the existing isomorphisms between the two classes of systems [10]. Among the various time-invariant representations, the *cyclic representation* is the most useful for our purposes.

This article aims to provide an unrestricted decoupling solution for discrete-time linear periodic systems using state feedback, including the requirement for stability. While it seems likely that for square systems the result can be deduced from [8] or [9], a direct argument appears more appropriate here because it presents a natural construction of a *nonregular* state feedback control law needed to decouple nonsquare systems with more inputs than outputs.

The recent results [6] are applied to the time-invariant cyclic representation of a periodic system. The first contribution offers an alternative to the findings of [8] and [9] when focused on square discrete-time systems and regular state feedback; this alternative relies on the Hermite normal form of the cyclic transfer function of the periodic system. The second, and more significant contribution, is the construction of nonregular decoupling state feedback for nonsquare discrete-time linear periodic systems with more inputs than outputs. This is thought to be the first solution to the problem.

## II. Periodic Systems

Consider a linear periodic discrete-time system:

$$x(t+1) = A(t)x(t) + B(t)u(t), \quad y(t) = C(t)x(t) + D(t)u(t) \qquad (1)$$

where $t \in \mathrm{Z}$, $u(t) \in \mathrm{R}^m$ is the input, $x(t) \in \mathrm{R}^n$ is the state, $y(t) \in \mathrm{R}^p$ is the output, and $A(\cdot), B(\cdot), C(\cdot)$, and $D(\cdot)$ are real periodic matrices of period $T$. We say that the system is *square* if $m = p$, and *nonsquare* otherwise; we are particularly interested in the case where $m > p$.

The *state transition matrix* of (1) is

$$\Phi(t,\tau) := \begin{cases} A(t-1)A(t-2)\cdots A(\tau), & t > \tau \\ I, & t = \tau. \end{cases}$$

The transition matrix over one period:

$$\Psi(\tau) := \Phi(\tau + T, \tau)$$

is called the *monodromy matrix* of (1) at time $\tau$.

We note that the dimension, $n$, of the state space of (1) may periodically change over time. In such cases, the matrix $A(t)$ is no longer square and its size is $n(t+1) \times n(t)$. Of course, $\Psi(\tau)$ is an $n(\tau) \times n(\tau)$ square matrix. Denoting $n_m$ its minimum size, the spectrum of $\Psi(\tau)$ is the union of a set of $n(\tau) - n_m$ zero eigenvalues and another set of $n_m$ eigenvalues, known as its *core spectrum*, which is independent of τ [10, Remark 3.1].

The system (1) is considered *stable* if, for any initial state, the free motion asymptotically approaches zero. By periodicity, system (1) is stable if and only if all the eigenvalues of $\Psi(\tau)$ are less than 1 in modulus [10, Proposition 3.3].

The output of the system (1) can be expressed as

This work was co-sponsored by the European Union under Project ROBOPROX CZ.02.01.01/00/ 22_008/0004590.

V. Kučera is with the Czech Institute of Informatics, Robotics, and Cybernetics, Czech Technical University in Prague, Czech Republic (e-mail: kucera@cvut.cz).

$$y(t)=M_0(t)u(t)+M_1(t)u(t-1)+M_2(t)u(t-2)+\cdots \tag{2}$$

where $M_i(t),\ i=0,1,...$ are known as the *Markov parameters*, which completely capture the input-output behavior of the system. Obviously, $M_i(t+T)=M_i(t)$ for each $t$ and $i = 0, 1, \ldots$

The periodicity of the Markov parameters suggests that the output response of the system at a generic time instant $t = kT + \tau$, with $\tau = 0, 1, \ldots, T-1$, can be expressed as a finite sum of the output responses of $T$ time-invariant systems indexed by the integer $\tau$ and possessing the Markov parameters $M_i(\tau), M_{i+T}(\tau), M_{i+2T}(\tau), \ldots$. For each $i = 0, 1, \ldots, T-1$ and each $\tau = 0, 1, \ldots, T-1$ we define as

$$H_i(z^T,\tau) := M_i(\tau)+M_{i+T}(\tau)z^{-T}+M_{i+2T}(\tau)z^{-2T}+\cdots \tag{3}$$

the transfer function from the sampled input $\hat{u}(k):=u(kT+\tau-i)$ to the sampled output $\hat{y}(k):=y(kT+\tau-i)$ and refer to it as the *sampled transfer function* of the system (1).

Referring to (1), the Markov parameters can be expressed as

$$\begin{aligned}
&M_0(t)=D(t),\\
&M_1(t)=C(t)B(t-1),\\
&M_2(t)=C(t)A(t-1)B(t-2),\\
&M_3(t)=C(t)A(t-1)A(t-2)B(t-3),\ \ldots
\end{aligned}$$

so that (3) implies

$$H_0(z^T,\tau)=D(\tau)+C(\tau)\left[z^T I-\Psi(\tau)\right]^{-1}\Phi(\tau,\tau-T+1)B(\tau)$$

$$H_i(z^T,\tau)=z^T C(\tau)\left[z^T I-\Psi(\tau)\right]^{-1}\Phi(\tau,\tau-i+1)B(\tau-i),\quad i=1,2,\ \ldots\ T-1.$$

With reference to a generic discrete-time signal $v(t)$ we introduce the *one-step-ahead shift operator* $\zeta$ defined by $\zeta(v(t)) := v(t+1)$. Then expression (2) can be rewritten in the form $y(t)=P(\zeta,t)u(t)$, where the *periodic transfer operator* $P(\zeta, t)$ is

$$\begin{aligned}
P(\zeta,t) &:= M_0(t)+M_1(t)\zeta^{-1}+M_2(t)\zeta^{-2}+\ldots\\
&= H_0(\zeta^T,t)+H_1(\zeta^T,t)\zeta^{-1}+\cdots+H_{T-1}(\zeta^T,t)\zeta^{-(T-1)}
\end{aligned} \tag{4}$$

for all $t\in \mathrm{Z}$, see [10, Section 1.2, p. 64]. Obviously, $P_i(t+T)=P_i(t)$ for each $t$ and $i = 0, 1, \ldots$

## III. Cyclic representation

Among the time-invariant representations of discrete-time periodic systems, the *cyclic representation* is the most practical for our purposes.

The operation of cycling a signal $v(t)\in \mathrm{R}^q,\ t\in \mathrm{Z}$, can be described as follows. Given an initial time instant $\tau$, define an augmented signal $\bar{v}_\tau(t)\in \mathrm{R}^{qT},\ t\in \mathrm{Z}$, as

$$\bar{v}_\tau(t):=\begin{bmatrix}\bar{v}_1(t)\\ \bar{v}_2(t)\\ \vdots\\ \bar{v}_T(t)\end{bmatrix}$$

where

$$\bar{v}_i(t):=\begin{cases} v(t), & t=kT+\tau+i-1,\ i=1,2,...,T\\ 0, & \text{otherwise.}\end{cases}$$

In this way, the vector $\bar{v}_\tau(t)$ has a unique nonzero subvector at each time point, cyclically shifting along the column blocks.

The cyclic representation at $\tau$ of the periodic system (1) is just the state-space relation among the cycled state, cycled input, and cycled output signals:

$$\bar{x}_\tau(t+1)=\bar{A}_\tau\bar{x}_\tau(t)+\bar{B}_\tau\bar{u}_\tau(t),\quad \bar{y}_\tau(t)=\bar{C}_\tau\bar{x}_\tau(t)+\bar{D}_\tau\bar{u}_\tau(t) \tag{5}$$

where

$$\bar{A}_\tau=\begin{bmatrix}
0 & 0 & \cdots & 0 & A(\tau+T-1)\\
A(\tau) & 0 & \cdots & 0 & 0\\
0 & A(\tau+1) & \cdots & 0 & 0\\
\cdots & \cdots & \ddots & \cdots & \cdots\\
0 & 0 & \cdots & A(\tau+T-2) & 0
\end{bmatrix}$$

$$\bar{B}_\tau=\begin{bmatrix}
0 & 0 & \cdots & 0 & B(\tau+T-1)\\
B(\tau) & 0 & \cdots & 0 & 0\\
0 & B(\tau+1) & \cdots & 0 & 0\\
\cdots & \cdots & \ddots & \cdots & \cdots\\
0 & 0 & \cdots & B(\tau+T-2) & 0
\end{bmatrix}$$

$$\bar{C}_\tau=\begin{bmatrix}
C(\tau) & 0 & \cdots & 0\\
0 & C(\tau+1) & \cdots & 0\\
\cdots & \cdots & \ddots & \cdots\\
0 & 0 & \cdots & C(\tau+T-1)
\end{bmatrix}$$

$$\bar{D}_\tau=\begin{bmatrix}
D(\tau) & 0 & \cdots & 0\\
0 & D(\tau+1) & \cdots & 0\\
\cdots & \cdots & \ddots & \cdots\\
0 & 0 & \cdots & D(\tau+T-1)
\end{bmatrix}.$$

The cyclic state-space representation has a unique structure. The eigenvalues of $\bar{A}_\tau$ are determined by the $T$th root of the eigenvalues of $\Psi(\tau)$ so the periodic system (1) is stable if and only if its cyclic representation is also stable.

The transfer function from $\bar{u}_\tau$ to $\bar{y}_\tau$ will be denoted as

$$\bar{W}_\tau(z):=\bar{C}_\tau(zI-\bar{A}_\tau)^{-1}\bar{B}_\tau+\bar{D}_\tau$$

and will be referred to as the *cyclic transfer function* of (1). It is a $pT \times mT$ matrix, and it inherits a particular block structure:

$$\bar{W}_\tau(z)=\begin{bmatrix}
H_0(z^T,\tau) & H_{T-1}(z^T,\tau)z^{-T+1} & \cdots & H_1(z^T,\tau)z^{-1}\\
H_1(z^T,\tau+1)z^{-1} & H_0(z^T,\tau+1) & \cdots & H_2(z^T,\tau+1)z^{-2}\\
\cdots & \cdots & \ddots & \cdots\\
H_{T-1}(z^T,\tau+T-1)z^{-T+1} & H_{T-2}(z^T,\tau+T-1)z^{-T+2} & \cdots & H_0(z^T,\tau+T-1)
\end{bmatrix}. \tag{6}$$

In particular, $\bar{W}_\tau(z)$ is block diagonal at $z = \infty$. Furthermore, it straightforwardly changes with time – the passing from $\bar{W}_\tau(z)$ to $\bar{W}_{\tau+1}(z)$ amounts to a permutation of the blocks [10, Section 6.3.3].

The zero structure (finite and infinite) of $\bar{W}_\tau(z)$ does not change with the initial time instant $\tau$. Therefore, we can define the *cyclic zero structure* of a periodic system as the zero structure of any associated cyclic representation.

## IV. The Cyclic Hermite Normal Form

The cyclic transfer function of any stable periodic system (1) features entries that are *proper and stable rational* functions in $z$. We will review some algebraic properties of these matrices.

The set of all proper and stable rational functions is a Euclidean domain, denoted as S. The Euclidean norm function is the *order* of nonzero elements. The order of a nonzero element $r = b/a$ of S, where $a$ and $b$ are polynomials, is defined as ord $r := \deg a - \deg b + \deg e$, where $e$ is the unique monic polynomial of greatest degree whose zeros lie outside the stability region $\Lambda := \{z \mid |z| < 1\}$ and which divides $b$. The units of S can now be described as the elements of order zero.

An element $p$ of S is said to be an *associate* of an element $q$ of S if $p = qu$ for some unit $u$ of S. This defines an equivalence relation over S, which decomposes S into equivalence classes with respect to this relation. The class that contains 0 consists of 0 alone, and the class containing 1 consists of the units. A set of elements of S, one from each equivalence class, is said to be a *complete set of nonassociates*. We choose this set to consist of 0 and the elements in the *normal form* $e / z^k$, where $k \geq 0$ is the order of the element and $e$ is any monic polynomial the zeros of which lie outside $\Lambda$. We denote this set as N.

Let $r$ be any nonzero element of S. We say that *$p$ is congruent to $q$ modulo $r$* if $r$ divides $p-q$. Once again, this is an equivalence relation over S. A set of elements of S, one from each equivalence class, is referred to as a *complete set of residues modulo r*. To complete the definition, a complete set of residues modulo 0 will be taken to be the entire set S, and $p$ is congruent to $q$ modulo 0 means that $p = q$. If $r$ is a nonzero element of N, we select this residue set to be either the zero element if $r$ is a unit, or else the elements of the form $f / z^l$, where $f$ is a polynomial and $l <$ ord $r$.

Matrices over S are referred to as *proper and stable rational matrices*. The set of square matrices over S forms a ring, and the units of this ring are known as *unimodular* matrices. A square matrix $U$ is unimodular if and only if it is nonsingular and its determinant is a unit of S. We define a $p \times m$ matrix $U$ as *row (or column) unimodular* if a greatest common divisor of all the $p \times p$ (or $m \times m$) minor determinants of $U$ is a unit of S, which means that $U$ can be completed to an unimodular matrix.

A square matrix $P$ of S with entries $p_{ij}$ is considered *lower triangular* if $p_{ij} = 0$ whenever $i < j$, and *diagonal* if $p_{ij} = 0$ whenever $i \neq j$. This terminology will also apply to *rectangular* matrices.

Let $P$ and $Q$ be $p \times m$ matrices over S. We say that $P$ is right equivalent to $Q$ if $P = QU$ for an $m \times m$ unimodular matrix $U$ over S. The canonical form for matrices over S, under one-sided unimodular transformations, is the *Hermite normal form* over S.

Every $p \times m$ matrix $P$ over S is right-equivalent to a lower triangular matrix, where each diagonal entry lies in N and the entries to the left of each diagonal entry belong to the complete set of residues modulo the diagonal entry, see [11, pp.15 and 18]. The matrix described above is the (column, lower triangular) *Hermite normal form* of $P$ over S.

The unimodular transformation can be achieved through the following elementary column operations on $P$: (i) interchanging two columns, (ii) multiplying a column by a unit of S, and (iii) adding $q$ times one column to another column, where $q$ belongs to S. Elementary row operations are defined in an analogous manner.

When $P$ is a cyclic transfer function, say $\overline{W}_\tau(z)$, we want to maintain the specific block structure of the matrix described in (6). To achieve this, we transform each $p \times m$ diagonal block of $\overline{W}_\tau(z)$ into the Hermite normal form over S by applying elementary operations to the respective columns of $\overline{W}_\tau(z)$. Finally, using elementary column operations (iii) alone, we replace as many off-diagonal entries as possible in all blocks with their residues modulo the diagonal entry in the corresponding row from a different block. The resulting matrix is referred to as the *cyclic Hermite normal form* over S of $\overline{W}_\tau(z)$. The unimodular transformation matrix, say $\overline{U}_\tau(z)$, is block diagonal at infinity.

## V. Problem Formulation

Our objective is to decouple a discrete-time periodic system of the form (1) using periodic state feedback:

$$u(t) = F(t)x(t) + G(t)v(t) \tag{7}$$

where $v(t) \in \mathrm{R}^p$ is an external input vector and $F(\cdot)$, $G(\cdot)$ are real periodic matrices of period $T$.

If $m = p$ and $G(t)$ is nonsingular for all $t$, the state feedback (7) is termed *regular*; otherwise, it is termed *nonregular*. Note that if $m > p$, decoupling can only be achieved using nonregular state feedback, whereas if $m < p$, only block decoupling is theoretically possible.

Authorizing nonregular state feedback, we do not attempt to stabilize and decouple the system simultaneously. Rather, we try to decouple it while maintaining its stability. If the system is not stable at the outset, it must first be stabilized and then *stability-preserving decoupling feedback* must be designed. We cannot destroy the existence of a decoupling state feedback when applying stabilizing state feedback. On the other hand, the application of *nonregular* decoupling feedback may destroy the system's stabilizability.

*The (row-by-row, or diagonal) Decoupling Problem*: Given a stable periodic system (1) with a periodic transfer operator $P(\zeta, t)$ that has full row-rank $p$, find a periodic state feedback control law (7) so that the closed-loop system

$$\begin{aligned} x(t+1) &= (A+BF)(t)x(t) + (BG)(t)v(t) \\ y(t) &= (C+DF)(t)x(t) + (DG)(t)v(t) \end{aligned} \tag{8}$$

is stable and its periodic transfer operator is diagonal and nonsingular for all $t$. □

In other words, the system is decoupled if for each $i = 1, 2, \ldots, p$, the $i$th component $y_i(\cdot)$ of $y(\cdot)$ in (8) is not influenced by the $j$th component $v_j(\cdot)$ of $v(\cdot)$ for all $j \neq i$. This indicates that all the Markov parameters of the closed-loop system are diagonal for all $t$.

The assumption rank $P(\zeta, t) = p$ is an admissibility condition imposed to prevent the decoupled system from having a diagonal yet *singular* periodic transfer operator.

## VI. Square Systems

Let us first consider the systems having as many inputs as outputs, that is, $m = p$ in (1). This assumption greatly simplifies the solution to the Decoupling Problem.

*Theorem 1*. Consider a stable system (1) with $m = p$, with the periodic transfer operator $P(\zeta, t)$ of rank $p$, and with the cyclic transfer function $\overline{W}_\tau(z)$. Then the Decoupling Problem admits a solution if and only if all the $p \times p$ blocks of the cyclic Hermite normal form over S of $\overline{W}_\tau(z)$ are diagonal.

*Proof*. To demonstrate the necessity, let $\overline{W}_\tau(z)$ be the cyclic transfer function of a stable system (1) with rank $P(\zeta, t) = p$. Let (7) be stability-preserving and decoupling state feedback that results in the closed-loop system (8) and the related cyclic transfer function $\overline{W}_\tau^{F,G}(z)$. The assumption rank $P(\zeta, t) = p$ indicates that $\overline{W}_\tau(z)$ is a nonsingular matrix.

The state feedback (7) also has a cyclic representation:

$$\overline{u}_\tau(t) = \overline{F}_\tau \overline{x}_\tau(t) + \overline{G}_\tau \overline{v}_\tau(t)$$

where

$$\overline{F}_\tau = \begin{bmatrix} F(\tau) & 0 & \cdots & 0 \\ 0 & F(\tau+1) & \cdots & 0 \\ \cdots & \cdots & \ddots & \cdots \\ 0 & 0 & \cdots & F(\tau+T-1) \end{bmatrix}$$

$$\bar{G}_\tau = \begin{bmatrix} G(\tau) & 0 & \cdots & 0 \\ 0 & G(\tau+1) & \cdots & 0 \\ \cdots & \cdots & \ddots & \cdots \\ 0 & 0 & \cdots & G(\tau+T-1) \end{bmatrix}.$$

Then

$$\bar{W}_\tau^{F,G}(z) = \bar{W}_\tau(z) U_\tau(z),$$

where

$$\bar{U}_\tau(z) := \left[ I - \bar{F}_\tau \bar{S}_\tau(z) \right]^{-1} \bar{G}_\tau \tag{9}$$

is the transfer function from $\bar{v}_\tau$ to $\bar{u}_\tau$, called the *compensator*, and where

$$\bar{S}_\tau(z) := (zI - \bar{A}_\tau)^{-1} \bar{B}_\tau \tag{10}$$

denotes the transfer function of (1) from the input $\bar{u}_\tau$ to the state $\bar{x}_\tau$.

Then, because the closed-loop system (8) is decoupled, all the Markov parameters of (8) are diagonal for all $t$, and (4) implies that all the blocks of $\bar{W}_\tau^{F,G}(z)$ are diagonal.

The key observation is that $\bar{U}_\tau(z)$ in (9) is a matrix over S that is *unimodular*. Therefore, $\bar{W}_\tau^{F,G}(z)$ is right equivalent to $\bar{W}_\tau(z)$ over S. This implies that $\bar{W}_\tau^{F,G}(z)$ is nonsingular. Transform $\bar{W}_\tau^{F,G}(z)$ into the cyclic Hermite normal form over S; this must also be the cyclic Hermite normal form over S of $\bar{W}_\tau(z)$. Note that elementary operation (i) is not utilized in this process, and operations (ii) and (iii) maintain the diagonal structure of $\bar{W}_\tau^{F,G}(z)$. Thus, the cyclic Hermite normal form of $\bar{W}_\tau(z)$ over S has all its blocks in diagonal form. This proves the necessity.

To prove the sufficiency, consider a stable system (1) with rank $P(\zeta, t) = p$ and with the cyclic transfer function $\bar{W}_\tau(z)$. Apply an unimodular transformation $\bar{U}_\tau(z)$ to bring $\bar{W}_\tau(z)$ to its cyclic Hermite normal form over S, denoted as $\bar{\Delta}_\tau(z)$, :

$$\bar{W}_\tau(z) \bar{U}_\tau(z) = \bar{\Delta}_\tau(z).$$

The matrix $\bar{\Delta}_\tau(z)$ inherits the block structure shown in (6) making it the transfer function of the cyclic representation of a periodic system. Additionally, all its blocks are assumed to be diagonal. Based on (4), the periodic transfer operator of this system is diagonal.

This means that the system is decoupled provided its periodic transfer operator has full row-rank $p$. The original system is assumed to have this property so $\bar{W}_\tau(z)$ is nonsingular. Since unimodular transformations do not alter the rank of the matrix, $\bar{\Delta}_\tau(z)$ is also nonsingular, and the property is preserved.

It remains to demonstrate that the action of the compensator $\bar{U}_\tau(z)$ on $\bar{W}_\tau(z)$ can be achieved by applying periodic state feedback of the form (7) to the given periodic system (1). To this end, note that $\bar{U}_\tau^{-1}(\infty)$ is block-diagonal and $\bar{U}_\tau^{-1}(z) - \bar{U}_\tau^{-1}(\infty)$ is strictly proper. By construction, the matrix $\bar{U}_\tau^{-1}(z) - \bar{U}_\tau^{-1}(\infty)$ and the input-to-state cyclic transfer function (10) share the same structure in the sense that the equation

$$\bar{U}_\tau^{-1}(z) - \bar{U}_\tau^{-1}(\infty) = \bar{K}\bar{S}_\tau(z)$$

admits a block-diagonal constant solution matrix $\bar{K} \in \mathrm{R}^{pT \times nT}$. Let

$$\bar{F}_\tau := -\bar{U}_\tau(\infty)\bar{K}, \quad \bar{G}_\tau := \bar{U}_\tau(\infty).$$

Then (9) is satisfied. The matrix $\bar{G}_\tau$ is block diagonal so that $\bar{F}_\tau$ is block diagonal as well. The block entries of $\bar{F}_\tau$ and $\bar{G}_\tau$ yield the matrices $F(t)$ and $G(t)$, which provide a decoupling control law.

The decoupled closed-loop system (8) has the cyclic transfer function $\bar{W}_\tau^{F,G}(z) = \bar{\Delta}_\tau(z)$. The system is stable since the monodromy matrices of (8) and (1) share the same unreachable spectrum. □

Note that the decoupling feedback constructed in the proof is *regular* as $G(t)$ is square and nonsingular for all $t$.

## VII. Nonsquare Systems

Suppose the system has more inputs than outputs, meaning $m > p$ in (1). If the cyclic Hermite normal form of the system does not have all its blocks diagonal, the redundant inputs may be utilized to decouple the system due to hidden dynamics. These dynamics can be linked to the input terminals using nonregular state feedback to modify the system's infinite and unstable zeros while relinquishing the extra input terminals.

The solution will modify the solution presented in [6] for time-invariant systems. Explicitly, we bring the cyclic representation (5) of the periodic system (1) to a specific form, called the *decoupling standard form*. We then define the *cyclic interactor* of the system, determine the e xistence of certain lists of nonnegative integers, and finally construct a nonregular decoupling state-feedback control law.

Finite lists of nonnegative integers play a crucial role in decoupling theory. Consider a list $\lambda := (\lambda_1, \lambda_2, ..., \lambda_l)$. We call $l$ the *length* of the list and denote it $\operatorname{sum}\lambda := \lambda_1 + \lambda_2 + ... + \lambda_l$. The *index set* $J_\lambda$ of $\lambda$ is defined as $J_\lambda := \{ i \mid \lambda_i \neq 0,\ i = 1, 2,\ ...,\ l \}$.

Consider two lists of nonnegative integers, $\lambda$ and $\mu$, of equal length. We say that $\mu$ is a s*ublist* of $\lambda$, and write $\mu \le \lambda$, if $\mu_i \le \lambda_i$ for all $i$. The sum $\lambda + \mu$ is defined elementwise. Provided that $\mu \le \lambda$, the difference $\lambda - \mu$ is also defined elementwise.

Consider two lists of nonnegative integers, $\lambda$ and $\mu$. Denote $\lambda'$ and $\mu'$ the lists obtained from $\lambda$ and $\mu$ by including or deleting zero elements so that they have equal length, denoted as $l$, and then arranging the lists in a nondecreasing order. We then say that $\lambda$ *dominates* $\mu$, and write $\lambda \succ \mu$, if

$$\lambda_1' + \lambda_2' + ... + \lambda_l' \ \ge\ \mu_1' + \mu_2' + ... + \mu_l', \quad i = 1, 2, ..., l.$$

The decoupling standard form can be obtained as follows:

(1) Let $\bar{W}_\tau(z)$ be the cyclic transfer function of a stable periodic system (1), satisfying the admissibility condition rank $P(\zeta, t) = p$, and bring $\bar{W}_\tau(z)$ to its cyclic Hermite normal form $\bar{\Delta}_\tau(z)$ over S by applying an unimodular transformation $\bar{U}_\tau(z)$.

(2) Consider the diagonal entries of all the $p \times m$ blocks of $\bar{\Delta}_\tau(z)$. Let $d_{i,j} / z^{-\delta_{i,j}}$ be the highest-order diagonal entry in column $j = 1, 2, \ldots, p$ of each $pT \times m$ column-block $i = 1, 2, \ldots, T$. We call

$$d_i := (d_{i,1}, d_{i,2}, ..., d_{i,p}), \quad i = 1, 2, ..., T \tag{11}$$

the list of the *first decoupling polynomials* and

$$\delta_i := (\delta_{i,1}, \delta_{i,2}, ..., \delta_{i,p}), \quad i = 1, 2, ..., T \tag{12}$$

the list of the *first decoupling orders* of the system.

(3) We determine the block-diagonal matrices $\bar{F}_\tau$ and $\bar{G}_\tau$ that satisfy (9) to implement the operation of $\bar{U}_\tau(z)$ on $\bar{W}_\tau(z)$ through the periodic state feedback expressed in (7) applied to given periodic system (1).

(4) The transformed system is maximally *unobservable*; we apply a block-diagonal state-transformation matrix $\bar{Q}_\tau$ to convert it to the standard form for unobservable systems [12, Sect. 3.4.A.2] to isolate the completely unobservable part of the system.

(5) Let the *reachability indices* of the completely unobservable subsystem be

$$\sigma_i := (\sigma_{i,1}, \sigma_{i,2}, ..., \sigma_{i,m-p}), \quad i = 1, 2, ..., T. \tag{13}$$

Using further state feedback and coordinate transformations, we bring this subsystem to its normal form [12, Sect. 3.4.D.1], featuring

$(m-p)T$ chains of pure delays with lengths equal to the reachability indices.

These chains are either *free* (having a single input) or *coupled* (also influenced by inputs to the observable part of the system so that only an initial subchain of length $\sigma^f_{i,j} < \sigma_{i,j}$ is free).

Next, we define the cyclic interactor:

(6) Let $\bar{\Delta}^*_\tau(z)$ be the $pT \times pT$ matrix obtained from $\bar{\Delta}_\tau(z)$ by removing the $m-p$ columns on the right of each column-block and annulling the diagonal entries of all the blocks except for the highest-order entries $d_{i,j} / z^{-\delta_{i,j}}$; the off-diagonal entries remain unchanged. The matrix is nonsingular based on the assumption that rank $P(\zeta, t) = p$.

(7) The inverse matrix of $\bar{\Delta}^*_\tau(z)$, denoted $\bar{\Phi}_\tau(z)$, is called the *cyclic interactor* of the system; it has only infinite and unstable poles. Let $f_{i,j}$ be the monic least common denominator of column $j$ in the $pT \times p$ block-column $i$ of $\bar{\Phi}_\tau(z)$, and let $\varphi_{i,j}$ denote the degree of $f_{i,j}$ increased by the greatest infinite-pole multiplicity occurring in column $j$ of the block-column $i$. We call

$$f_i := (f_{i,1}, f_{i,2}, \ldots, f_{i,p}),\ i = 1,2,\ldots,T \tag{14}$$

the list of the *second decoupling polynomials* and

$$\varphi_i := (\varphi_{i,1}, \varphi_{i,2}, \ldots, \varphi_{i,p}),\ i = 1,2,\ldots,T \tag{15}$$

the list of the *second decoupling orders* of the system.

We are now prepared to present the solvability result for the Decoupling Problem:

*Theorem 2*: The Decoupling Problem has a solution if and only if there are lists of nonnegative integers

$$\varepsilon_i := (\varepsilon_{i,1}, \varepsilon_{i,2}, \ldots, \varepsilon_{i,p}),\ i = 1,2,\ldots,T \tag{16}$$

with each element of the lists being a multiple of $T$,

$$\eta_i := (\eta_{i,1}, \eta_{i,2}, \ldots, \eta_{i,m-p}),\ i = 1,2,\ldots,T \tag{17}$$

and

$$\eta^*_i := (\eta^*_{i,1}, \eta^*_{i,2}, \ldots, \eta^*_{i,p}),\ i = 1,2,\ldots,T \tag{18}$$

the latter being a selection of the elements of $\eta_i$ satisfying $J_{\eta^*_i} = J_{\varepsilon_i}$ for $i = 1, 2, \ldots, T$, such that the following conditions are satisfied for each $i = 1, 2, \ldots, T$:

$$\delta_i + \varepsilon_i \geq \varphi_i$$

$$\delta_i + \varepsilon_i \geq \eta^*_i$$

$$\eta_i \leq \sigma_i$$

$$\eta_i \succ \omega_i$$

$$\text{sum}\,\varepsilon_i = \text{sum}\,\eta_i = \text{sum}\,\omega_i$$

where

$$\omega_i := (\omega_{i,1}, \omega_{i,2}, \ldots, \omega_{i,p}),\ i = 1,2,\ldots,T \tag{19}$$

is the list of the orders of the invariant factors of the matrix

$$\bar{\Phi}_\tau(z)\,\text{diag}(f_{i,j} / z^{\delta_{i,j}+\varepsilon_{i,j}}),\ i = 1,2,\ldots,T \text{ and } j = 1,2,\ldots,p. \tag{20}$$

*Proof:* The proof involves a slight refinement of the result [6, Theorem 1] applicable to time-invariant systems. This is feasible due to a time-invariant representation, specifically the cyclic one, of the periodic system. For the decoupled system to remain periodic, all decoupling operations must maintain the specific structure of the cyclic transfer functions. In particular, state feedback along with input and state coordinate transformations must be block-diagonal. Consequently, the corresponding unimodular transformations must also be block diagonal at infinity.

The cyclic Hermite normal form of the decoupled system's cyclic transfer function is multi-diagonal, featuring T diagonals each of length T, cut to form super-diagonals and sub-diagonals. This indicates a specific structure of the interactor, which shows how the unobservable chains of pure delays should be connected to decouple the system. To maintain periodicity, each connected chain must also be periodic with length T. □

Because the system to be decoupled has a full-row rank periodic transfer operator for all $t$, the matrix $\bar{\Delta}^*_\tau(z)$ and the interactor $\bar{\Phi}_\tau(z)$ are unique. This implies that the lists (11), (12), (13), (14), (15), and (19), which condition the solvability of the Decoupling Problem, are *invariant* with respect to state feedback and input and state coordinate transformations.

Theorem 2 indicates that the system can be decoupled using *regular* state feedback, with the lists of zeros in (16), if and only if the first and the second decoupling orders are equal.

If the conditions of Theorem 2 are met, the process for constructing a *nonregular* decoupling state feedback is as follows:

(8) The key step is to identify the lists (16), (17), and (18). The lists (16) specify the required increase in decoupling orders to achieve decoupling; the lists (17) outline which parts of the unobservable pure delay chains should be connected to the input terminals to facilitate this increase; and the lists (18) indicate which of these chains will be linked to the external input terminals. By connecting the chains of pure delays, we raise the decoupling orders $\varepsilon_i$ and establish the missing unstable invariant zeros $e_i := f_i / d_i$ specified by the interactor.

(9) We begin by connecting free subchains. For each $i = 1, 2, \ldots, T$, we set $\varepsilon_i := \varphi_i - \delta_i$ whenever each element of the list is a multiple of $T$; otherwise, we adjust the element to the nearest multiple of $T$. We choose $\eta_i$ such that $\eta_i \leq \sigma^f_i$ and $\text{sum}\,\eta_i = \text{sum}\,\varepsilon_i$. We calculate $\omega_i$ from (20) and verify whether $\eta_i \succ \omega_i$. We select $\eta^*_i$ such that $J_{\eta^*_i} = J_{\varepsilon_i}$ and $\varepsilon_i + \delta_i \geq \eta^*_i$.

(10) If no such lists exist, we repeat this procedure for the system modified by connecting a selected set of coupled chains to a chosen set of input terminals This creates new couplings, so we bring the modified system to the decoupling standard form once again, yielding modified invariants, and continue connecting the remaining free subchains until the necessary lists are found.

(11) Denote $Z_1(z)$ the $pT \times pT$ matrix defined in (20), which is already part of the decoupling compensator. To realize the compensator through state feedback using the available unobservable chains, we make $Z_1(z)$ into an $(m-p)T \times (m-p)T$ matrix, denoted $V_{22}(z)$, having the same lists of nonunit invariant factor orders $\omega_i$, $i = 1, 2, \ldots, T$ but with column orders specified by the lists $\eta_i$, $i = 1, 2, \ldots, T$.

For simplicity, let us assume that $p = m - p$. Then

$$V_{22}(z) Z_2(z) = -V_{21}(z) Z_1(z) \tag{21}$$

for *unimodular* matrices $V_{21}(z)$ and $Z_2(z)$. The modifications for $p > m - p$ and $p < m - p$ are described in [6, Sect. XVII]. Then

$$\begin{bmatrix} 0 & Z_2^{-1}(z) \\ V_{21}(z) & V_{22}(z) \end{bmatrix} \begin{bmatrix} Z_1(z) \\ Z_2(z) \end{bmatrix} = \begin{bmatrix} I \\ 0 \end{bmatrix}$$

where the square matrix, denoted as $V(z)$, is $mT \times mT$ and *unimodular*. Reducing the column orders of $Z_2^{-1}(z)$ below $\eta_i$, $i = 1, 2, \ldots, T$ by performing elementary row operations on $V(z)$ results in a matrix identity of the form

$$\begin{bmatrix} V_{11}(z) & V_{12}(z) \\ V_{21}(z) & V_{22}(z) \end{bmatrix} \begin{bmatrix} Z_1(z) \\ Z_2(z) \end{bmatrix} = \begin{bmatrix} I \\ 0 \end{bmatrix}. \tag{22}$$

(12) We now move the rows and columns of $V(z)$ defined by the submatrix $V_{22}(z)$ to their original block positions. Specifically, for $i$ = 1, 2, …, $T$, we relocate the columns of $V(z)$ defined by the $i$th block of $m - p$ columns of $V_{22}(z)$ to the end of the $i$th block of $p$ columns of $V(z)$. We apply the same procedure for the rows of $V(z)$, effectively arranging the matrix into blocks of size $m \times m$. We adjust the rows of the $Z(s)$-matrix in (22) accordingly. We denote the structured form of (22) as

$$\bar{V}(z)\bar{Z}(z) = \bar{L}. \tag{23}$$

Note that $\bar{L}$ is block-diagonal; elementary operations can also be applied to make $\bar{V}(\infty)$ and $\bar{Z}(\infty)$ block-diagonal as well.

(13) The matrix $\bar{Z}(z) = \bar{V}^{-1}(z)\bar{L}$ is a decoupling compensator that is column unimodular and realizable through nonregular state feedback applied to (1). We follow the construction outlined in the proof of Theorem 1 to obtain the matrices

$$\bar{F} = -\bar{V}^{-1}(\infty)\bar{K}, \quad \bar{G} = \bar{V}^{-1}(\infty)\bar{L},$$

which provides a solution to the Decoupling Problem. The monodromy matrix of the decoupled system has all eigenvalues equal to zero; if the system is reachable, we can change its core spectrum at will.

## VIII. Examples

In the following examples, we examine discrete-time periodic systems (1) with a period of $T = 2$ and an initial sampling time $\tau = 0$.

*Example 1:* Let

$$A(0) = \begin{bmatrix} 2 & 0 \\ 0 & 1 \end{bmatrix}, \quad A(1) = \begin{bmatrix} 1 & 0 \\ 1 & 1 \end{bmatrix}$$

$$B(0) = \begin{bmatrix} 1 & 1 \\ 0 & 1 \end{bmatrix}, \quad B(1) = \begin{bmatrix} 0 & 1 \\ 1 & 1 \end{bmatrix}$$

and $C(0) = C(1) = I$, $D(0) = D(1) = 0$. The system is not stable, but it is stabilizable, sfor example with the feedback gains

$$F_s(0) = \begin{bmatrix} -1 & 1 \\ 0 & -1 \end{bmatrix}, \quad F_s(1) = \begin{bmatrix} 0 & 0 \\ -1 & 0 \end{bmatrix}$$

and with $G_s(0) = G_s(1) = I$, yielding

$$\begin{gathered} (A + BF_s)(0) = \begin{bmatrix} 1 & 0 \\ 0 & 0 \end{bmatrix}, \quad (A + BF_s)(1) = \begin{bmatrix} 0 & 0 \\ 0 & 1 \end{bmatrix} \\ (BG_s)(0) = \begin{bmatrix} 1 & 1 \\ 0 & 1 \end{bmatrix}, \quad (BG_s)(1) = \begin{bmatrix} 0 & 1 \\ 1 & 1 \end{bmatrix}. \end{gathered} \tag{24}$$

The cyclic transfer function of (24) is:

$$\bar{W}(z) = \left[\begin{array}{cc|cc} 0 & 0 & 0 & z^{-1} \\ 0 & z^{-2} & z^{-1} & z^{-1} \\ \hline z^{-1} & z^{-1} & 0 & z^{-2} \\ 0 & z^{-1} & 0 & 0 \end{array}\right]$$

and it is also the cyclic input-to-state transfer function. Applying the unimodular transformation

$$\bar{U}(z) = \left[\begin{array}{cc|cc} 1 & -1 & -z^{-1} & 0 \\ 0 & 1 & 0 & 0 \\ \hline 0 & -z^{-1} & -1 & 1 \\ 0 & 0 & 1 & 0 \end{array}\right]$$

brings (24) into the Hermite normal form over S:

$$\bar{\Delta}(z) = \bar{W}(z)\bar{U}(z) = \left[\begin{array}{cc|cc} 0 & 0 & z^{-1} & 0 \\ 0 & 0 & 0 & z^{-1} \\ \hline z^{-1} & 0 & 0 & 0 \\ 0 & z^{-1} & 0 & 0 \end{array}\right].$$

Theorem 1 implies that the system can be decoupled by regular state feedback. The equation

$$\bar{U}^{-1}(z) - \bar{U}^{-1}(\infty) = \bar{K}\bar{S}(z)$$

is satisfied by

$$\bar{K} = \begin{bmatrix} 1 & 0 & 0 & 0 \\ 0 & 0 & 0 & 0 \\ 0 & 0 & 0 & 0 \\ 0 & 0 & 0 & 1 \end{bmatrix}$$

so that

$$\bar{F}_d = \bar{U}(\infty)\bar{K} = \left[\begin{array}{cc|cc} -1 & 0 & 0 & 0 \\ 0 & 0 & 0 & 0 \\ \hline 0 & 0 & 0 & -1 \\ 0 & 0 & 0 & 0 \end{array}\right], \quad \bar{G}_d = \bar{U}(\infty) = \left[\begin{array}{cc|cc} 1 & -1 & 0 & 0 \\ 0 & 1 & 0 & 0 \\ \hline 0 & 0 & -1 & 1 \\ 0 & 0 & 1 & 0 \end{array}\right]$$

is the cyclic representation of a periodic decoupling control law (7) for the stabilized system (24). Hence,

$$F_d(0) = \begin{bmatrix} -1 & 0 \\ 0 & 0 \end{bmatrix}, \quad F_d(1) = \begin{bmatrix} 0 & -1 \\ 0 & 0 \end{bmatrix}$$

$$G_d(0) = \begin{bmatrix} 1 & -1 \\ 0 & 1 \end{bmatrix}, \quad G_d(1) = \begin{bmatrix} -1 & 1 \\ 1 & 0 \end{bmatrix}$$

and the decoupled system is actually time-invariant,

$$\begin{gathered} A_d(0) = A_d(1) = 0, \quad B_d(0) = B_d(1) = I, \\ C_d(0) = C_d(1) = I, \quad D_d(0) = D_d(1) = 0, \end{gathered}$$

giving rise to the periodic transfer operator

$$P^{F,G}(\zeta, t) = \begin{bmatrix} \zeta^{-1} & 0 \\ 0 & \zeta^{-1} \end{bmatrix}$$

for all values of $t$. □

*Example 2:* Given the stable system (1) with the matrices

$$\begin{gathered} A(0) = \begin{bmatrix} 0 & 0 & 0 \\ 0 & 0 & 0 \end{bmatrix}, \quad A(1) = \begin{bmatrix} 1 & 0 \\ 0 & 1 \\ 0 & 0 \end{bmatrix} \\ B(0) = \begin{bmatrix} 0 & 1 & 0 \\ 0 & 0 & 1 \end{bmatrix}, \quad B(1) = \begin{bmatrix} 0 & 0 & 0 \\ 0 & 0 & 1 \\ 0 & 0 & 1 \end{bmatrix} \\ C(0) = \begin{bmatrix} 0 & 0 & 0 \\ -4 & 0 & 1 \end{bmatrix}, \quad C(1) = \begin{bmatrix} 0 & 0 \\ 0 & 0 \end{bmatrix} \\ D(0) = \begin{bmatrix} 1 & 0 & 0 \\ 1 & 1 & 0 \end{bmatrix}, \quad D(1) = \begin{bmatrix} 1 & 0 & 0 \\ 0 & 1 & 0 \end{bmatrix}, \end{gathered} \tag{25}$$

determine decoupling state feedback.

The input-to-state cyclic transfer function of system (25) is

$$\bar{S}(z) = \left[\begin{array}{ccc|ccc} 0 & z^{-2} & 0 & 0 & 0 & z^{-1} \\ 0 & 0 & z^{-2} & 0 & 0 & z^{-1} \\ 0 & 0 & 0 & 0 & 0 & 0 \\ \hline 0 & z^{-1} & 0 & 0 & 0 & 0 \\ 0 & 0 & z^{-1} & 0 & 0 & 0 \end{array}\right]$$

and the resulting cyclic transfer function

$$\overline{W}(z)=\left[\begin{array}{ccc|ccc} 1 & 0 & 0 & 0 & 0 & 0 \\ 1 & 1-4z^{-2} & 0 & 0 & 0 & z^{-1} \\ \hline 0 & 0 & 0 & 1 & 0 & 0 \\ 0 & 0 & 0 & 0 & 1 & 0 \end{array}\right]$$

is already in Hermite normal form over S. The system has unstable cyclic zeros at +2 and −2.

The cyclic interactor is

$$\overline{\Phi}(z)=\left[\begin{array}{cc|cc} 1 & 0 & 0 & 0 \\ -\frac{z^2}{z^2-4} & \frac{z^2}{z^2-4} & 0 & 0 \\ \hline 0 & 0 & 1 & 0 \\ 0 & 0 & 0 & 1 \end{array}\right]$$

and the decoupling invariants are

$$\delta_1=(0,2),\delta_2=(0,0);\quad \varphi_1=(2,2),\varphi_2=(0,0)$$
$$d_1=(1,z^2-4),d_2=(1,1);\quad f_1=(z^2-4,z^2-4),f_2=(1,1).$$

The system (25) is in the decoupling standard form with

$$\sigma_1=(2),\ \sigma_2=(0).$$

The unobservable chain is free. The lists of indices

$$\varepsilon_1=(2,0),\ \varepsilon_2=(0,0);\quad \eta_1=(2),\ \eta_2=(0);\quad \eta_1^*=(2,0),\ \eta_2^*=(0,0)$$

with the matrix

$$Z_1(z)=\left[\begin{array}{cc|cc} 1-4z^{-2} & 0 & 0 & 0 \\ -1 & 1 & 0 & 0 \\ \hline 0 & 0 & 1 & 0 \\ 0 & 0 & 0 & 1 \end{array}\right]$$

and the lists of its invariant factor orders $\omega_1=(2,0),\omega_2=(0,0)$ satisfy the conditions of Theorem 2, so the Decoupling Problem admits a solution.

Since $m=3$ and $p=2$, we calculate

$$V_{22}(z)=\left[\begin{array}{c|c} 1-4z^{-2} & 0 \\ \hline 0 & 1 \end{array}\right]$$

and the row-unimodular matrices

$$V_{21}(z)=\begin{bmatrix} -1 & 0 & 0 & 0 \\ 0 & -1 & 0 & 0 \end{bmatrix},\quad Z_2(z)=\begin{bmatrix} 1 & 0 & 0 & 0 \\ 0 & 1 & 0 & 0 \end{bmatrix}$$

from (21). The matrix identity (23) is as follows:

$$\left[\begin{array}{ccc|ccc} 0 & 0 & 1 & 0 & 0 & 0 \\ 0 & 1 & 1 & 0 & 0 & 0 \\ -1 & 0 & 1-4z^{-2} & 0 & 0 & 0 \\ \hline 0 & 0 & 0 & 1 & 0 & 0 \\ 0 & 0 & 0 & 0 & 1 & 0 \\ 0 & 0 & 0 & 0 & 0 & 1 \end{array}\right]\left[\begin{array}{cc|cc} 1-4z^{-2} & 0 & 0 & 0 \\ -1 & 1 & 0 & 0 \\ 1 & 0 & 0 & 0 \\ \hline 0 & 0 & 1 & 0 \\ 0 & 0 & 0 & 1 \\ 0 & 0 & 0 & 0 \end{array}\right]=\left[\begin{array}{cc|cc} 1 & 0 & 0 & 0 \\ 0 & 1 & 0 & 0 \\ 0 & 0 & 0 & 0 \\ \hline 0 & 0 & 1 & 0 \\ 0 & 0 & 0 & 1 \\ 0 & 0 & 0 & 0 \end{array}\right].$$

The equation

$$\overline{V}(z)-\overline{V}(\infty)=\overline{K}\overline{S}(z)$$

is satisfied with

$$\overline{K}=\left[\begin{array}{ccc|cc} 0 & 0 & 0 & 0 & 0 \\ 0 & 0 & 0 & 0 & 0 \\ 0 & -4 & 4 & 0 & 0 \\ \hline 0 & 0 & 0 & 0 & 0 \\ 0 & 0 & 0 & 0 & 0 \\ 0 & 0 & 0 & 0 & 0 \end{array}\right]$$

so that the decoupling compensator

$$\overline{Z}(z)=\left[\begin{array}{cc|cc} 1-4z^{-2} & 0 & 0 & 0 \\ -1 & 1 & 0 & 0 \\ 1 & 0 & 0 & 0 \\ \hline 0 & 0 & 1 & 0 \\ 0 & 0 & 0 & 1 \\ 0 & 0 & 0 & 0 \end{array}\right]$$

is realizable through nonregular state feedback in cyclic form

$$\overline{F}=-\overline{V}^{-1}(\infty)\overline{K},\quad \overline{G}=\overline{V}^{-1}(\infty)\overline{L}.$$

Explicitly,

$$F(0)=\begin{bmatrix} 0 & -4 & 4 \\ 0 & 0 & 0 \\ 0 & 0 & 0 \end{bmatrix},\quad F(1)=\begin{bmatrix} 0 & 0 \\ 0 & 0 \\ 0 & 0 \end{bmatrix}$$

$$G(0)=\begin{bmatrix} 1 & 0 \\ -1 & 1 \\ 1 & 0 \end{bmatrix},\quad G(1)=\begin{bmatrix} 1 & 0 \\ 0 & 1 \\ 0 & 0 \end{bmatrix}$$

and

$$(A+BF)(0)=\begin{bmatrix} 0 & 0 & 0 \\ 0 & 0 & 0 \end{bmatrix},\quad (A+BF)(1)=\begin{bmatrix} 1 & 0 \\ 0 & 1 \\ 0 & 0 \end{bmatrix}$$

$$(BG)(0)=\begin{bmatrix} -1 & 1 \\ 1 & 0 \end{bmatrix},\quad (BG)(1)=\begin{bmatrix} 0 & 0 \\ 0 & 0 \\ 0 & 0 \end{bmatrix}$$

$$(C+DF)(0)=\begin{bmatrix} 0 & -4 & 4 \\ -4 & -4 & 5 \end{bmatrix},\quad (C+DF)(1)=\begin{bmatrix} 0 & 0 \\ 0 & 0 \end{bmatrix}$$

$$(DG)(0)=\begin{bmatrix} 1 & 0 \\ 0 & 1 \end{bmatrix},\quad (DG)(1)=\begin{bmatrix} 1 & 0 \\ 0 & 1 \end{bmatrix}.$$

The periodic transfer operator of the closed-loop system is

$$P^{F,G}(\zeta,0)=\begin{bmatrix} 1-4\zeta^{-2} & 0 \\ 0 & 1-4\zeta^{-2} \end{bmatrix},\quad P^{F,G}(\zeta,1)=\begin{bmatrix} 1 & 0 \\ 0 & 1 \end{bmatrix}.\qquad \square$$

*Example 3:* Consider a stable system (1) described by

$$A(0)=\begin{bmatrix} 0 & 0 \\ 0 & 0 \\ 0 & 1 \end{bmatrix},\quad A(1)=\begin{bmatrix} 1 & 0 & 0 \\ 0 & 0 & 0 \end{bmatrix}$$

$$B(0)=\begin{bmatrix} 0 & 1 & 0 \\ 0 & 0 & 1 \\ 1 & 0 & 0 \end{bmatrix},\quad B(1)=\begin{bmatrix} 1 & 0 & 0 \\ 0 & 0 & 1 \end{bmatrix}$$

$$C(0)=\begin{bmatrix} 0 & 0 \\ 1 & 0 \end{bmatrix},\quad C(1)=\begin{bmatrix} 0 & 0 & 0 \\ 1 & 0 & 0 \end{bmatrix} \tag{26}$$

$$D(0)=\begin{bmatrix} 1 & 0 & 0 \\ 0 & 0 & 0 \end{bmatrix},\quad D(1)=\begin{bmatrix} 1 & 0 & 0 \\ 0 & 1 & 0 \end{bmatrix}.$$

The objective is to determine periodic decoupling state feedback.

The input-to-state cyclic transfer function is as follows:

$$\overline{S}(z)=\left[\begin{array}{ccc|ccc} 0 & z^{-2} & 0 & z^{-1} & 0 & 0 \\ 0 & 0 & 0 & 0 & 0 & z^{-1} \\ \hline 0 & z^{-1} & 0 & 0 & 0 & 0 \\ 0 & 0 & z^{-1} & 0 & 0 & 0 \\ z^{-1} & 0 & 0 & 0 & 0 & z^{-2} \end{array}\right].$$

The cyclic transfer function of the system

$$\overline{W}(z)=\left[\begin{array}{ccc|ccc} 1 & 0 & 0 & 0 & 0 & 0 \\ 0 & z^{-2} & 0 & z^{-1} & 0 & 0 \\ \hline 0 & 0 & 0 & 1 & 0 & 0 \\ 0 & z^{-1} & 0 & 0 & 1 & 0 \end{array}\right]$$

is already in Hermite normal form over S.

The cyclic interactor is

$$\overline{\Phi}(z)=\left[\begin{array}{cc|cc} 1 & 0 & 0 & 0 \\ 0 & z^{2} & -z & 0 \\ \hline 0 & 0 & 1 & 0 \\ 0 & 0 & 0 & 1 \end{array}\right]$$

and the decoupling invariants are

$$\delta_1=(0,2),\delta_2=(0,0);\quad \varphi_1=(0,2),\varphi_2=(1,0)$$

$$d_1=(1,1),d_2=(1,1);\quad f_1=(1,1),f_2=(1,1).$$

The system (26) is in decoupling standard form with

$$\sigma_1=(1),\ \sigma_2=(2).$$

One unobservable chain is free, while the other is coupled, and we have

$$\sigma_1^f=(1),\ \sigma_2^f=(1).$$

Only chains of length $T=2$ can be connected, so we select

$$\varepsilon_1=(0,0),\ \varepsilon_2=(2,0).$$

We take the other lists as

$$\eta_1=(1),\ \eta_2=(1);\quad \eta_1^*=(0,0),\ \eta_2^*=(1,0).$$

Matrix

$$Z_1(z)=\left[\begin{array}{cc|cc} 1 & 0 & 0 & 0 \\ 0 & 1 & -z^{-1} & 0 \\ \hline 0 & 0 & z^{-2} & 0 \\ 0 & 0 & 0 & 1 \end{array}\right]$$

has the orders of its invariant factors $\omega_1=(0,0),\omega_2=(0,2).$ Since the conditions of Theorem 2 are satisfied, a (nonregular) decoupling state feedback exists.

Following the design procedure, we calculate the matrix

$$V_{22}(z)=\left[\begin{array}{c|c} 1 & -z^{-1} \\ \hline -z^{-1} & 0 \end{array}\right]$$

having the same nonunit invariant factor orders as $Z_1(z),$ but column orders $\eta_1=(1),\ \eta_2=(1).$ We obtain the row unimodular matrices

$$V_{21}(z)=\begin{bmatrix} 0 & -1 & 0 & 0 \\ 0 & z^{-1} & 1 & 0 \end{bmatrix},\quad Z_2(z)=\begin{bmatrix} 0 & 1 & 0 & 0 \\ 0 & 0 & 1 & 0 \end{bmatrix}$$

from (21) so that (23) has the form

$$\left[\begin{array}{ccc|ccc} 1 & 0 & 0 & 0 & 0 & 0 \\ 0 & 0 & 1 & 0 & 0 & 0 \\ 0 & -1 & 1 & 0 & 0 & -z^{-1} \\ \hline 0 & 0 & 0 & 0 & 0 & 1 \\ 0 & 0 & 0 & 0 & 1 & 0 \\ 0 & z^{-1} & -z^{-1} & 1 & 0 & 0 \end{array}\right]\left[\begin{array}{cc|cc} 1 & 0 & 0 & 0 \\ 0 & 1 & -z^{-1} & 0 \\ 0 & 1 & 0 & 0 \\ \hline 0 & 0 & z^{-2} & 0 \\ 0 & 0 & 0 & 1 \\ 0 & 0 & 1 & 0 \end{array}\right]=\left[\begin{array}{cc|cc} 1 & 0 & 0 & 0 \\ 0 & 1 & 0 & 0 \\ 0 & 0 & 0 & 0 \\ \hline 0 & 0 & 1 & 0 \\ 0 & 0 & 0 & 1 \\ 0 & 0 & 0 & 0 \end{array}\right].$$

Periodic decoupling state feedback (7) directly results from the diagonal blocks of the matrices

$$\overline{F}=\left[\begin{array}{cc|ccc} 0 & 0 & 0 & 0 & 0 \\ 0 & -1 & 0 & 0 & 0 \\ 0 & 0 & 0 & 0 & 0 \\ \hline 0 & 0 & -1 & 1 & 0 \\ 0 & 0 & 0 & 0 & 0 \\ 0 & 0 & 0 & 1 & 0 \end{array}\right],\quad \overline{G}=\left[\begin{array}{cc|cc} 1 & 0 & 0 & 0 \\ 0 & 1 & 0 & 0 \\ 0 & 1 & 0 & 0 \\ \hline 0 & 0 & 0 & 0 \\ 0 & 0 & 0 & 1 \\ 0 & 0 & 1 & 0 \end{array}\right].$$

The cyclic representation of the decoupled system is

$$\overline{W}^{F,G}(z)=\left[\begin{array}{cc|cc} 1 & 0 & 0 & 0 \\ 0 & z^{-2} & 0 & 0 \\ \hline 0 & 0 & z^{-2} & 0 \\ 0 & z^{-1} & 0 & 1 \end{array}\right]$$

and the periodic transfer operator is diagonal for all time instants:

$$P^{F,G}(\zeta,0)=\begin{bmatrix} 1 & 0 \\ 0 & \zeta^{-2} \end{bmatrix},\quad P^{F,G}(\zeta,1)=\begin{bmatrix} \zeta^{-2} & 0 \\ 0 & 1+\zeta^{-1} \end{bmatrix}.$$

□